# The Effect of Disorder on a Quantum Phase Transition

J.F. DiTusa[1*], S. Guo[1], D.P. Young[1], R. T. Macaluso[2], D.A. Browne[1], N.L. Henderson[2], and J.Y. Chan[2]

[1] Department of Physics and Astronomy, Louisiana State University, Baton Rouge, Louisiana 70803, USA

[2] Department of Chemistry, Louisiana State University, Baton Rouge, Louisiana 70803, USA.

The conductivity and magnetization of $Fe_{1-x}Co_xS_2$ were measured to investigate quantum critical behavior in disordered itinerant magnets. Small $x$ (<0.001) is required to convert insulating iron pyrite into a metal, followed by a paramagnetic-to-ferromagnetic metal transition at $x = 0.032+/-0.004$. Singular contributions are discovered that are distinct from those at either metal-insulator or magnetic transitions. Our data reveal that disorder and low carrier density associated with proximity to a metal-insulator transition fundamentally modifies the critical behavior of the magnetic transition.

PACS Numbers 72.20.-i; 75.47.-m

[*]To whom correspondence should be addressed. (email ditusa@phys.lsu.edu)

The provocative hypothesis that metals cannot be described by conventional band structure theory when in proximity to magnetic instabilities has recently inspired many notable investigations[1-8]. Realizations of materials on the verge of magnetic ordering rely on the appearance of spontaneous magnetic moments as pressure is applied or small changes in chemical formulae are made. Such exquisite control of material parameters allows tuning of the critical temperature for ordering so that it approaches zero, thereby producing a quantum critical point (QCP) [9,10]. Many investigations focusing on clean crystalline metals have unambiguously clarify QCPs and successfully connect QCP behavior to the properties of rare earth compounds and transition metal oxides[1-5]. However, many of these more complicated systems, including the superconducting oxides of copper, are poor metals with small charge carrier concentrations and significant disorder[11-13]. As such, there is a second, and perhaps equally important, QCP nearby in their phase diagrams governing the metal-to-insulator (MI) transition. We report here on the discovery of novel charge transport properties in proximity to zero temperature MI and ferromagnetic (FM) to paramagnetic (PM) transitions that are not found at either magnetic or MI QCPs alone.

Disorder is generally thought to either destroy or broaden long range and long-lived fluctuations that provide many of the interesting properties of phase transitions. However, it is now well established that diffusive charge motion in disordered metals, resulting in poor screening of Coulomb interactions, leads to the appearance of singularities in the electronic density states and non-analytic temperature ($T$) and magnetic field ($H$) dependencies of transport and thermal properties[14]. In fact, the consequences of disorder and interactions between charge carriers near MI transitions have been the focus of investigation for decades[14,15]. Recent theoretical work has suggested that these weak localization effects couple strongly to the FM-PM quantum critical behavior since FM spin fluctuations in a 3-dimensional (3D) metal are constrained to diffuse in unison with charge fluctuations[8]. This fundamentally modifies the critical behavior of the FM-PM transition. With the purpose of exploring the role of disorder and the proximity of a MI transition to a magnetic QCP we have experimentally investigated the pyrite series $Fe_{1-x}Co_xS_2$. Here we observe qualitatively unique low $T$ transport properties in the vicinity of the zero $T$ phase transition between paramagnetism and itinerant FM ordering. The non-Fermi liquid behavior we measure has similarities to that seen in superconducting cuprates in high magnetic fields[12,13].

The parent compound of the series we have chosen to investigate is pyrite, $FeS_2$ commonly known as "fool's gold", in which we have exchanged periodic table neighbor Co for Fe to span the phase diagram of Fig. 1. This system is advantageous because the Co concentration can be varied over the entire range, $x=0$ to $x=1$, without a change in crystal structure, although some $CoS_2$ - $FeS_2$ phase separation occurs in the region around $x=0.5$[16,17]. Our samples are single crystals produced from high purity starting materials by standard iodine vapor techniques. We have characterized these crystals by performing X-ray diffraction and Energy dispersive X-ray microanalysis in order to establish the true Co concentration of our single crystals. The Hall effect data of Fig. 1c establish the sign (electrons) and density of the carriers added by Co substitution and demonstrates a systematic increase in carrier concentration with $x$. These data reveal the existence of a metallic state in the concentration range studied ($0.001 <\!/= x <\!/= 1$)[18].

In Figs. 1a and b we demonstrate the variations in the electrical conductivity ($\sigma$) and magnetic susceptibility ($\chi$) that occur on the Fe rich side of the phase diagram including the well-known PM insulating properties of $FeS_2$ on the far left side of the figure[16,19,20]. In contrast, CoS$_2$ is a classic itinerant FM with a Curie temperature ($T_c$) of 120 K[21] and a highly polarized (95%) electron gas at low $T$ [16,17,19,22]. For small Co concentrations ($x$<0.03) these crystals remain PM over the entire $T$ range of Fig. 1b while for $x$>0.03 we observe a shallow peak in χ at low $T$ signaling a magnetic ordering. Since the magnetic moment associated with the Curie-like $T$ dependence of χ above $T_c$ is 5 to 7 times larger than the saturated moment in the ordered phase below $T_c$ (Fig. 1c), the magnetism is largely itinerant[19,21-23]. An Arrott plot analysis[24] of the $T$ and $H$ dependent magnetization ($M$) clearly reveals the signature of a FM transition with $T_c$ increasing with $x$ (Fig. 1b) consistent with the peak in χ(T). Interpolations of the Arrott analysis to zero $T$ for samples with $x$<0.03 are not consistent with a finite $T$ phase transition. Furthermore, fits of power-law forms to the $T_c(x)$ data in Fig. 1b, including that expected for a FM QCP[15], indicate a zero $T$ critical point at a Co concentration above 0.03. We conclude that a FM to PM QCP is located in the region between our $x$=0.028 and our $x$=0.036 samples.

As the color conductivity plot of Fig. 1a points out, we find metallic behavior (σ > 100 Ω$^{-1}$ cm$^{-1}$, dσ/dT < 0) for all Co doped samples beginning with our lowest Co concentration crystal at $x$=0.001. One common measure of the degree of disorder in a metal is the value of $k_Fl$, where $k_F$ is the Fermi wave vector and $l$ is the mean free path. Our estimates of $k_Fl$ from the carrier densities of Fig. 1c and conductivities of Fig. 1a range from 2 to 15 for 0.001<$x$<0.17 confirming that these crystals are weakly metallic.

Perhaps the most intriguing feature of this phase diagram for our purposes, is the proximity of two zero $T$ phase transitions; the MI phase transition below $x$ = 0.001 and the PM to FM phase transition at $x$=0.032 +/- 0.004. Because this second transition follows closely behind in Co doping from the first, and because our $k_Fl$ values are small, the metallic state from which the ferromagnetism evolves is not that of a clean Fermi liquid. Instead, this metal is disordered and there remain significant correlations in the motion of the charge carriers. This is of special significance since it is the moments associated with these diffusively conducting charge carriers that align to create the itinerant magnetic state.

Our first indication that the metallic state created by Co substitution is not that of a simple Fermi liquid comes from the $T$ dependence of the resistivity ($\rho = 1/\sigma$). Fig. 2a demonstrates that $\rho$ is much more weakly $T$ dependent than a well-developed metallic behavior characterized by $\rho \propto T^3$ up to $T^5$ below the Debye $T$[25]. The $T$ dependence we observe is $\rho \propto T^{1.5+\backslash-0.2}$ which, when combined with a low $T$ contribution described in detail below, yields the solid lines in the figure. Such pseudo-linear $T$ dependence is well established as a marginal Fermi liquid (MFL) behavior closely associated with the zero $T$ magnetic phase[5,7,9,26]. However, the experimental error in the $T$ exponent too large to allow us to distinguish between a $T^{1.5}$ dependence and the $T^{5/3}$ dependence of $\rho$ reported in ref. 5. It is interesting to note that the extrapolation of a linear-in-$T$ form for $\rho$ has a zero $T$ intercept close to zero with a $\sigma$ slope between $-0.9$ and $-3.1$ $\Omega^{-1}\text{cm}^{-1}$, reminiscent of the behavior of the high $T_c$ superconducting cuprates[11].

Previous investigations of the QCP in clean ferromagnets ($\rho < 1\mu\Omega$ cm) report that this MFL behavior continues down to very low $T$ or is surprisingly interrupted by a superconducting transition[3-6]. In contrast, the proximity of the MI transition to the FM quantum critical point in $Fe_{1-x}Co_xS_2$ results in a novel contribution to $\sigma$ evident in the upturn in $\rho$ at $T$ below 20 K (note that from here on we use $\sigma$ ($= 1/\rho$) instead of $\rho$ since $\sigma$ is considered to be a more fundamental quantity.). This low $T$ contribution to $\sigma$ displayed in the inset to Fig. 2a and in Fig. 2b is observed to be strongly $T$ and $H$ dependent with power-law-like dependencies. As such, this singular σ resembles that found near simple PM MI transitions such as that in Si:P[15].

Several mechanisms are known to provide contributions to σ at low *T* including weak localization effects, the Kondo effect, spin glass formation, and paramagnon effects. A large number of investigations both experimental and theoretical have established that weak localization results in square-root dependence of σ in *H* and *T* for 3D systems near MI transitions[14,15,27]. However, the application of magnetic fields to our samples strongly *increases* the conductivity by as much as a 70% for fields of 9 T (Fig. 3a), behavior *opposite* in sign from these well-known electron-electron (e-e) interaction effects in disordered conductors. Fits of power laws to our raw data show a *T* and *H* dependence more consistent with $T^{0.3+/-0.1}$ (Fig. 2a inset) and $H^{0.8+/-0.1}$ than the square-root-like behavior of typical weak localization effects. The Kondo effect results from magnetic impurities in metals and provides a logarithmic-in-*T* correction to σ over an extended *T* range. Such T dependence does not represent our data well and the application of a magnetic field simply suppresses Kondo anomalies where as magnetic fields reverse the sign of the σ anomaly in $Fe_{1-x}Co_xS_2$[28]. Spin glass formation provides a zero field correction of the wrong sign[29] and simple paramagnon formation results in a $T^2$ dependence of σ[30].

A clue to the origins of the low *T* conductivity is in the systematic changes that occur with Co concentration as we parameterize in Fig. 1d by plotting $d\sigma/dT^{1/3}$ between 1.8 and 10 K at zero and 5 T. At *H*=0 this derivative is positive and increasing with *x* for *x*<0.01, but becomes small and negative at large values of *x*. At high fields $d\sigma/dT^{1/3}$ is negative throughout and has a minimum just beyond $x_c$. The zero and high field values of $d\sigma/dT^{1/3}$ vanish systematically as the MI transition is approached, where-as the high-field derivative has a maximum absolute value in the vicinity of the FM-PM QCP. This clearly connects the low $T$ anomaly in $\sigma$ with the critical behavior for ferromagnetism. For FM samples the singular $T$ dependence is cut-off below 0.5 K as shown for our $x$=0.05 sample in the inset to Fig. 2b, perhaps due to the appearance of a static $M$ in these samples.

The strong power-law-like increase in $\sigma$ with $H$ near $x_c$ is displayed in Fig. 3a. The FM regions of $T$ and $x$ are characterized by a sharp singular behavior at zero $H$ due to the spontaneous $M$. This is in contrast to the $H^2$ behavior at low fields in the PM regions. We have checked that the magnetoconductance (MC) is nearly independent of the direction of the current with respect to $H$ as pointed out in Fig. 3a. The similarity of the transverse and longitudinal MCs constrain the orbital contributions to the MC to be small, and therefore, similar to classic e-e interactions, the MC is dominated by a $H$ coupling to the carrier spins.

In the bottom half of Fig. 3 we demonstrate the field and $x$ dependence of $M$ measured at 1.8 K. The data clearly shows the continuous changes in $M$ as the ground state of the system varies from an enhanced paramagnet to a ferromagnet. Samples on the small $x$ side of $x_c$ display the typical behavior of PM materials with a linear $M(H)$ for fields between 0 and 0.2 T. As a static $M$ builds up for samples beyond $x$=0.036, $M$ becomes less $H$ dependent. Fits to the form $M=cH^p$ reveal a decrease in $p$ from 1 to ~1/4 as shown in the inset to Fig. 3b. Although there is significant scatter in the data, the value of $p$ near $x_c$ is consistent with that found in proximity to magnetic critical points in clean metallic magnets[2,5].

If, as we propose, the unusual $T$ and $H$ dependent $\sigma$ is a reflection of the critical behavior of the ordering spins, a scaling of the data in $H/T$ should be possible after allowing for *T* dependent corrections. Following the guidance of recent theoretical work[8] we test such scaling in Fig. 4. The suggested form has $\sigma \propto T^{1/3}$ with logarithmic corrections; that is $\sigma - \sigma_o = a[T/T_F\, g(\ln T/T_F)]^{1/3}$ where *a* is a nonuniversal constant, $T_F$ is the Fermi temperature, $g(y) = \Sigma\, \{[c(3)y]^n/n!\, \exp[-b\, n(n-1)]\}$, the sum is from $n = 0$ to infinity, $b = \frac{1}{2} \ln(3/2)$, and $c(3)$ is of order 1[8]. A fit to this form with $\sigma_o$, *a*, and $c(3)$ varied for the best representation to the data at *H*=0 between 1.8 and 10 K is shown by the solid line in the inset to Fig. 2a and in Fig. 2b for the *x*= 0.007 and 0.036 samples. The scaling of the data demonstrated in Fig. 4a for samples close to the transition is excellent. Attempts to scale the data to forms that excluded the logarithmic corrections of this type, including those found near classic MI transitions, proved to be unsatisfactory. For samples with $x>x_c$ (*x*=0.085 shown in the figure) we have used $H_{eff} = H + \alpha M$ as our modification of *H* as the spontaneous *M* becomes a significant factor in the effective magnetic field[27]. Here $\alpha$ is varied to determine the smallest difference of the scaled data to a piece-wise linear form ($\alpha = 710$). Furthermore, we show in the figure that in the low *H/T* limit the scaled data are consistent with an $(H/T)^2$ behavior while at $H/T >1$ an $(H/T)^{3/4}$ form is most appropriate. Figs. 2b and 3a include the simple asymptotes extracted from the scaling analysis of Fig. 4a for further comparison to the *x*=0.036 conductivity data. Although the model of Belitz *et al.*[8] appears consistent with our data, our analysis does not exclude different underlying physical explanations.

Fig. 4b demonstrates that the simplest scaling of our $\chi$ data holds near in the critical region. For samples with $x<x_c$ the data resemble the expected scaling form for free magnetic moments with an enhanced gyromagnetic ratio (g = 4)[25]. This is shown in the figure by the black line for the somewhat poorly scaled *x*=0.02 sample. In contrast, very near the critical concentration (*x*=0.028) the scaling is of high quality and $\chi T$ is described by a $\ln(H/T)$ dependence over a large range in $H/T$. Samples just beyond $x_c$ have a similar $H/T$ dependence except that the appearance of a spontaneous $M$ at the lowest $T$ destroys the scaling for small $H/T$.

We have presented here the first, to our knowledge, systematic exploration of the consequences of disorder to a magnetic QCP. As such, our data may have relevance to the long-standing mystery surrounding the normal state transport of the high $T_c$ cuprate superconductors. The linear T-dependent ρ of the cuprates was the first to be identified as a MFL behavior due to the proximity of the magnetic QCP[26]. Measurements of σ of these materials at $H$ large enough to suppress the superconductivity consistently revealed low $T$ logarithmic singularities[12,13]. The proximity to both a PM-FM and an MI quantum critical point in $Fe_{1-x}Co_xS_2$ yield weakly singular behavior of σ(T,H) at low $T$. We suggest that the low $T$ insulating-like behavior of the copper oxides may be a reflection of the proximity to two such critical points in much the same way. However, it is not clear that the disorder induced long-range spatial correlations of charge carriers has similar consequences on the critical behavior of the staggered $M$ of an antiferromagnet, or the randomly frozen moments of a spin glass.

JFD and JYC acknowledge the support of the National Science Foundation under contract DMR01-03892, DMR-0237664 and the ACS - Petroleum Research Fund. DPY acknowledges the support of the Louisiana Board of Regents under contract number 2001-04-RD-A-11 .

1. A. Yeh *et al.*, Nature **419**, 459 (2002).
2. A. Schroder *et al.*, Nature **407**, 351 (2000).
3. C. Pfleiderer *et al.*, Nature **412**, 58 (2001).
4. C. Pfleiderer, S. R. Julian, G.G. Lonzarich, Nature **414**, 427 (2001).

5. M. Nicklas *et al*., Phys. Rev. Lett. **82**, 4268 (1999).

6. S. S. Saxena *et al*., Nature **406**, 587 (2000).

7. A. J. Millis, Phys. Rev. B **48**, 7183 (1993).

8. D. Belitz, T. R. Kirkpatrick, R. Narayanan, T. Vojta, Phys. Rev. Lett. **85**, 4602 (2000).

9. J. A. Hertz, Phys. Rev. B **14**, 1165 (1976).

10. S. Sachdev, *Quantum Phase Transitions* (Cambridge Univ. Press, Cambridge, 1999).

11. H. Takagi *et al*., Phys. Rev. Lett**. 69**, 2975 (1992).

12. S. Ono *et al*., Phys. Rev. Lett. **85**, 638 (2000).

13. P. Fournier *et al*., Phys. Rev. Lett. **81**, 4720 (1998).

14. See for e.g. P. A. Lee, T. V. Ramakrishnan, Rev. Mod. Phys. **57**, 287 (1985).

15. T. F. Rosenbaum, *et al*., Phys. Rev. B **27**, 7509 (1983).

16. H. S. Jarrett *et al*., Phys. Rev. Lett. **21**, 617 (1968).

17. S. Ogawa, T. Teranishi, Phys. Lett. **42A**, 147 (1972).

18. All measurements were performed on single crystals of average size 2 mm x 1 mm x 1 mm. The electrical resistance was measured on irregularly shaped crystals with thin platinum wires attached with silver epoxy. We collected transverse and longitudinal magnetoresistance and Hall effect measurements at 17 Hz using lock-in techniques. Magnetization measurements were made in fields between −5 and 5 T in a SQUID magnetometer. Lattice parameters of all compositions in the $Fe_{1-x}Co_xS_2$ series (pyrite structure, $Pa\bar{3}$ spacegroup) were obtained from refinement of full data collections at 298 K on a single-crystal X-ray diffractometer.

19. G. L. Zhao, J. Callaway, M. Hayashibara, Phys. Rev. B **48**, 15781 (1993).

20. T. Harada, J. Phys. Soc. Jpn. **67,** 1352 (1998).

21. T. Moriya, *Spin Fluctuations in Itinerant Electron Magnetism* (ed. Fulde P.) (Springer, Berlin/Heidelberg/New York/Tokyo, 1985).

22. I. I. Mazin, Appl. Phys. Lett. **77**, 3000 (2000).

23. E. P. Wohlfarth, J. Mag. Mag. Mater. **7**, 113 (1978).

24. A. Arrott, J. E. Noakes, Phys. Rev. Lett. **19**, 786 (1967).

25. See, e.g., N. W. Ashcroft, & N. D. Mermin, *Solid State Physics* (Saunders, Philadelphia, 1976).

26. C. M. Varma, P. B. Littlewood, S. Schmittrink, E. Abrahams, A. E. Ruckenstein, Phys. Rev. Lett. **63**, 1996 (1989).

27. N. Manyala *et al.*, Nature **404**, 581 (2000).

28. M. D. Daybell, W.A. Steyert, Phys. Rev. Lett. **20**, 195 (1968).

29. K. Binder, A P. Young, *Rev. Mod. Phys.* **58**, 801 (1986).

30. A. Tari, & B. R. Coles, *J. Phys. F* **1**, L69 (1971).

**Figure Captions**

Fig. 1 **Phase diagram of $Fe_{1-x}Co_xS_2$. (A) Conductivity $\sigma$ versus temperature and Co concentration. (B) Magnetic susceptibility $\chi = M / H$ at 0.005 T versus temperature and Co concentration. Blue dots represent the Curie temperature determined from Arrott plots[24] of our field and temperature dependent magnetization data. Red dots are the Curie temperatures measured by Jarrett *et al*[16]. Blue and red lines represent the best fit of a $T_c \propto (x-x_c)^{3/4}$ form and a $T_c \propto (x-x_c)^{1/2}$ form to the $T_c(x)$**

data respectively. Blue and red lines represent fits of $(x-x_c)^{0.75}$ and $(x-x_c)^{0.5}$ dependencies to $T_c(x)$ indicating a quantum critical point near x=0.032. (C) Apparent carrier concentration $n_{app}$ as measured by the Hall effect at 300 K (filled blue circles), density of spin ½ paramagnetic moments as measured from the Curie-Weiss behavior above $T_c$ (light blue diamonds) and the concentration of saturated magnetic moments at 1.8 K and 5 T (green boxes) versus Co concentration *x*. Solid black line represents the density of Co atoms. (D) derivative of the conductivity with the cubed-root of temperature $d\sigma / dT^{1/3}$ at 4 K versus Co concentration parameterizing the low temperature contributions to $\sigma$. Dashed line in all sections of the figure denote the critical concentration for the appearance of the ferromagnetic phase (x=0.032+/-0.004).

Fig. 2 **Temperature dependence.** (A) The resistivity $\rho$ for crystalline $Fe_{1-x}Co_xS_2$ with x= 0.004 (dark blue asterisks), 0.005 (red diamonds), 0.007 (light blue triangles), 0.02 (blue-green squares), 0.036 (green Xs), and 0.05 (yellow filled circles). Solid black lines represent $T^{1.5}$ dependent resistivities added to the low temperature contributions described in the text. Dashed blue-green line represents a linear dependence separately for the x=0.007 sample. Inset The conductivity of our x=0.007 (light blue triangles) and 0.036 (green Xs) samples with the $T^{1.5}$ contributions subtracted. The lines are the best fits of $T^{1/2}$ form (dashed purple lines), a $T^{1/3}$ form (dashed blue lines), and the form suggested by Belitz *et al.*(*8*) described in the text (black line). (B) The conductivity $\sigma$ of $Fe_{0.964}Co_{0.036}S_2$ in fields of 0 (blue circles), 1 (light blue filled squares), 2 (green triangles), 3 (light green diamonds), 4 (yellow squares), and 5 T (orange filled circles). Solid black line represents a fit of the form described in the text to the *H*=0 data. Solid green line

**represents the same solid green line in figure 4a in order to display the H/T >1 fits at 5 T. Inset Change in the conductivity $\Delta\sigma=\sigma(T) - \sigma(0.07$ K$)$ versus temperature for $Fe_{0.95}Co_{0.05}S_2$ at temperatures below 2 K.**

Fig. 3 **Field dependence of magnetoconductivity and magnetization. (A) Transverse (*x*=0.036, *x*=0.05, and *x*=0.07) and longitudinal (*x*=0.07) magnetoconductivity of $Fe_{1-x}Co_xS_2$ at various temperatures. Dashed black and orange lines represent the fits to the scaled data in figure 4a for H/T < 1 (black) and H/T >1 (orange). (B) Magnetization *M* at 1.8 K for $Fe_{1-x}Co_xS_2$ for x=0.007 (purple circles), x=0.02 (filled blue squares), x=0.028 (dark green triangles), x=0.036 (light green diamonds) x=0.042 (yellow squares), x=0.05 (filled red circles), x=0.057 (filled blue triangles), x=0.07 (light blue x), x=0.085 (blue-green asterisks) and x=0.10 (filled orange diamonds). Inset Power law *p* for best fits of the form $cH^p$ to the *M* (*H*) data at 1.8 K for fields between 0.01 and 0.2 T. Purple line represents *p*=2/3 predicted for the quantum critical point by Belitz et al[8] and blue line *p*=3/4 found by Nicklas et al[5] in $Ni_xPd_{1-x}$.**

Fig. 4**, Scaling of the magnetoconductivity and magnetic susceptibility. (A) Scaling plot of the conductivity $(\sigma - \sigma_o) / (a\ [T/T_F\ g(\ln(T_F/T)]^{1/3})$ with $g(y) = \Sigma\ \{[c(3)y]^n/n!\ \exp[-b\ n(n-1)]\}$ where the sum is from *n* = 0 to infinity, *b* = ½ ln(3/2), *c*(3) is of order 1, and *a* is a constant[8] versus H/T; $\sigma_o$, *a*, and *c*(3), determined from fits of this form to the H=0 data for x=0.007 and x=0.036. For the x=0.085 sample we have plotted the same function versus $H_{eff}/T$ with $H_{eff}$ taken as $H + \alpha M$ and with $\sigma_o$ and $\alpha$ determined by the best scaling of all our *T* and *H* dependent data. Solid lines in the**

figure represent fits to the data of $(H/T)^2$ forms for $H/T<1$ and $(H/T)^{3/4}$ for $H/T>1$ for $x=0.036$ and $x=0.085$. Inset **The range of $H$ and $T$ covered.** (B) Scaling plot of the magnetic susceptibility, $\chi T$ versus $H/T$ for $x=0.02$, $x=0.028$, and $x=0.042$. Solid black line represents the form expected for non-interacting magnetic moments (Curie form) with a gyromagnetic factor of 4. Purple line is a fit of a logarithmic $H/T$ dependence for $x=0.028$ and the red line represents the same for $x=0.042$ data at 10 K only. Inset **The range of $T$ and $H$ covered.**

Figure 1

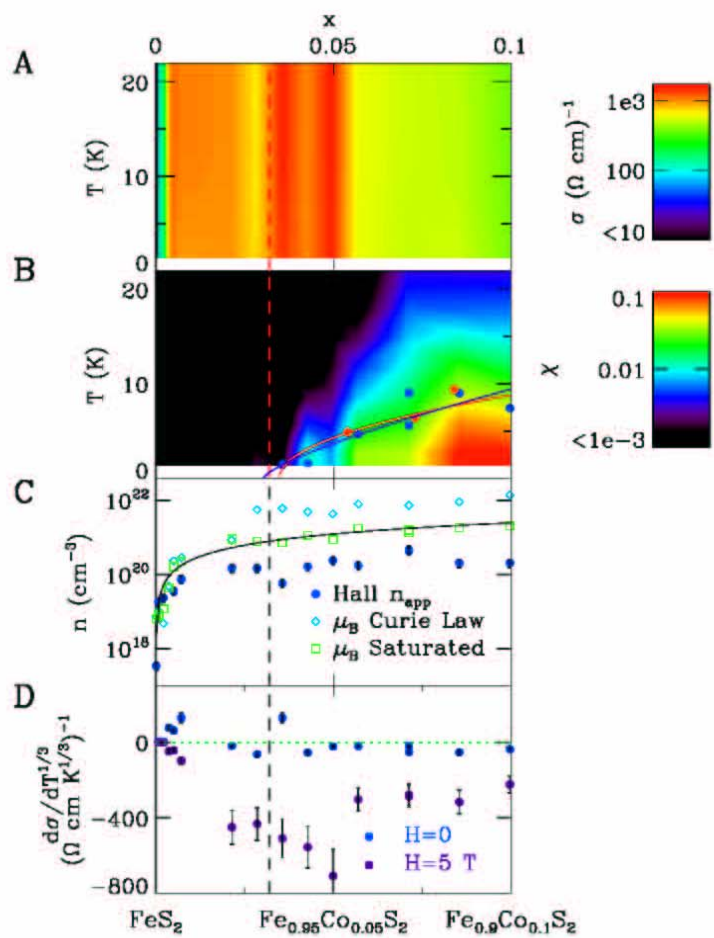

Figure 2

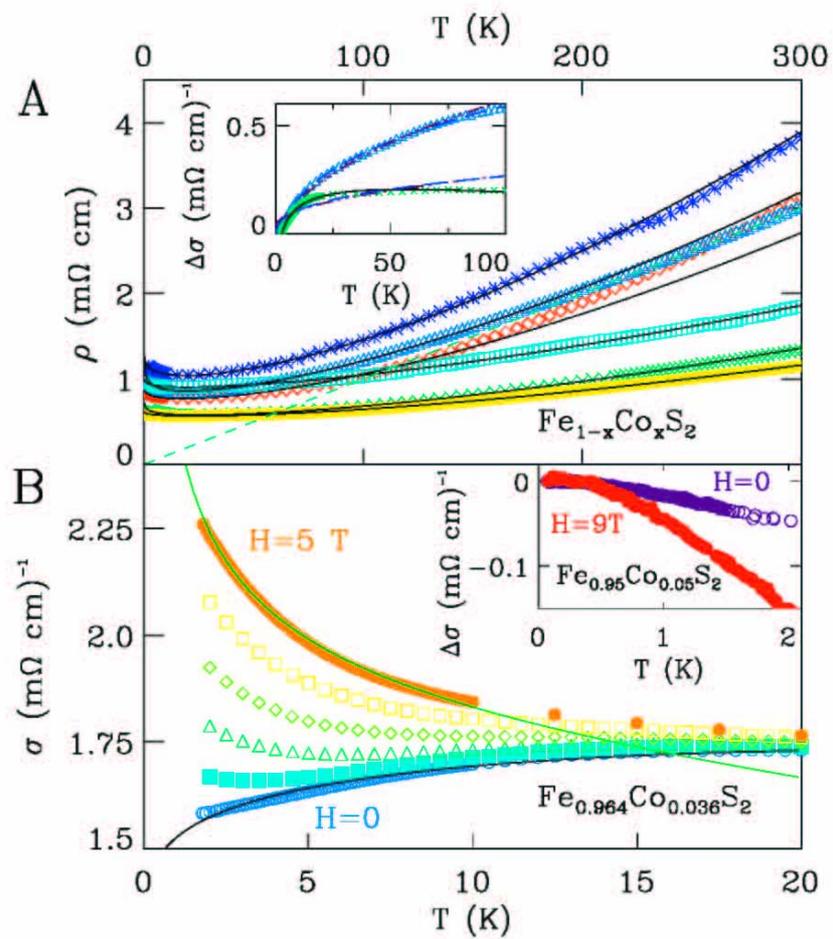

Figure 3

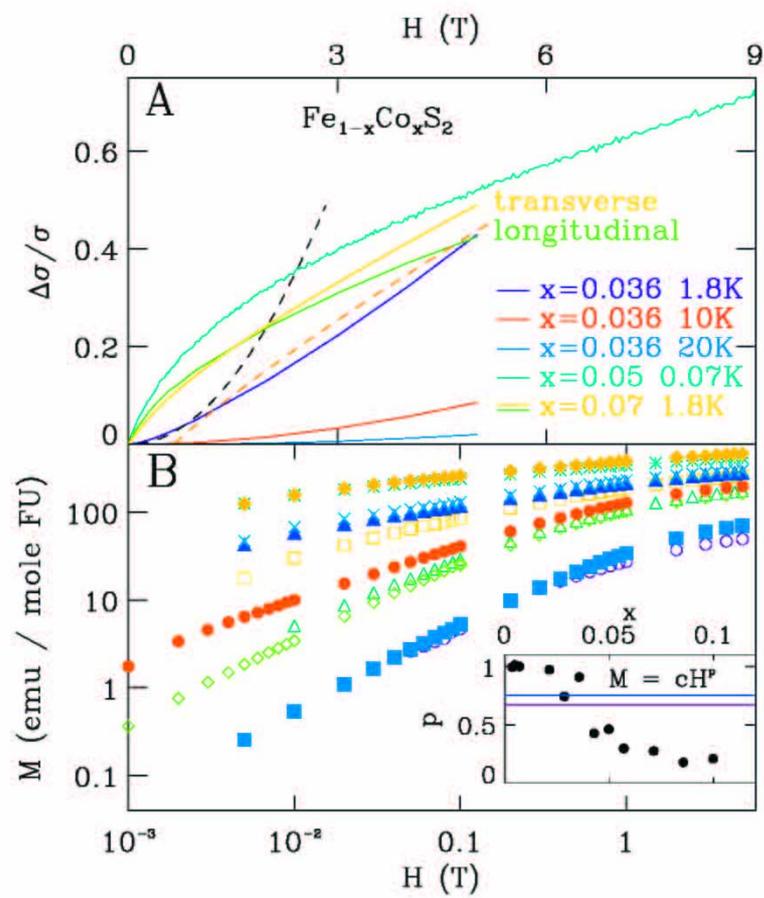

Figure 4

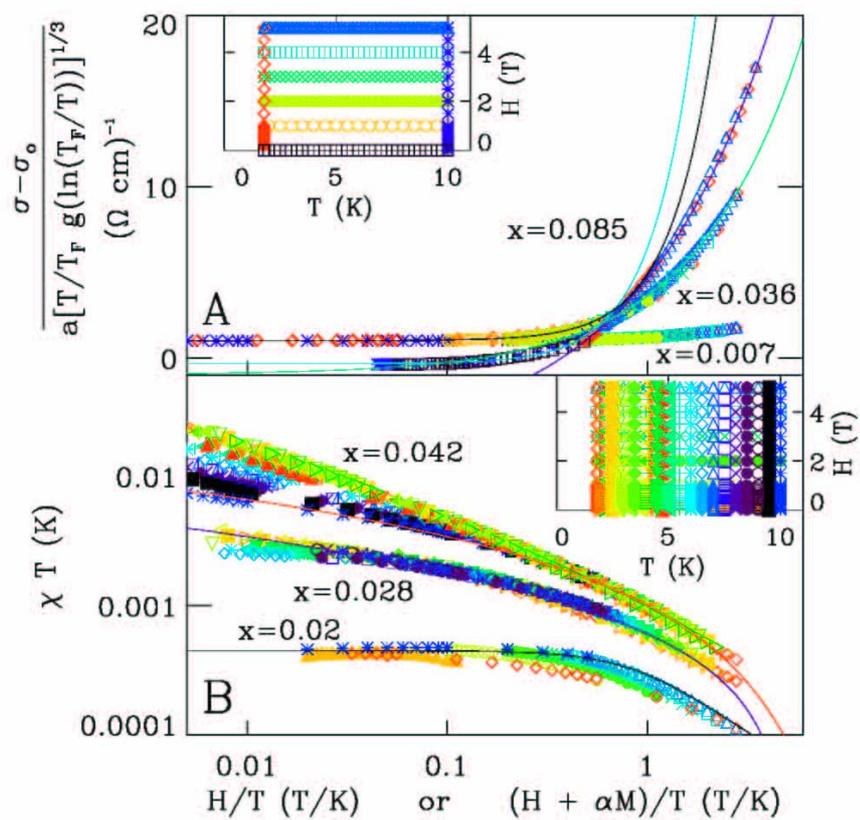